\documentclass[fleqn,10pt]{wlscirep}
\usepackage{multirow}

\title{Free energy of a chemotactic model with nonlinear diffusion}

\author[1,$\ast$]{Seung~Ki~Baek}
\author[2,$\dagger$]{Beom~Jun~Kim}

\affil[1]{Department of Physics, Pukyong National University, Busan 48513, Korea}
\affil[2]{Department of Physics, Sungkyunkwan University, Suwon 16419, Korea}

\affil[$\ast$]{seungki@pknu.ac.kr}
\affil[$\dagger$]{beomjun@skku.edu}

\begin{abstract}
The Patlak-Keller-Segel equation is a canonical model of chemotaxis to describe
self-organized aggregation of organisms interacting with chemical signals. We
investigate a variant of this model, assuming that the organisms exert effective
pressure proportional to the number density. From the resulting set of partial
differential equations, we derive a Lyapunov functional that can also be
regarded as the free energy of this model, and minimize it with a Monte Carlo
method to detect the condition for self-organized aggregation. Focusing on
radially symmetric solutions on a two-dimensional disc, we find that the
chemical interaction competes with diffusion so that aggregation occurs when the
relative interaction strength exceeds a certain threshold. Based on the analysis
of the free-energy landscape, we argue that the transition from a homogeneous
state to aggregation is abrupt yet continuous.
\end{abstract}

\begin{document}

\flushbottom
\maketitle
\thispagestyle{empty}

\section*{Introduction}
\label{sec:intro}

Ants communicate with each other through the use of pheromones to adjust their
collective behaviour~\cite{Holldobler1990,Vicsek2012,VelaPerez2015}.
This mechanism often leads to intriguing self-organized patterns.
For example, their foraging path can be understood as solving a
certain optimization problem in terms of time and energy
costs~\cite{Fewell1988,Denny2001,Couzin2003,Dussutour2004,Tao2004,Dussutour2006},
and the shape of the path is predictable by Fermat's principle of least
time~\cite{Goss1989,Reid2010,Oettler2013}.
From a biological point of view, especially in the context of natural selection,
it is highly plausible that an ant colony benefits from the ability of
organizing a foraging path.
It is also worth noting that the key
ingredient is not an individual ant with little computational capacity, but
the interaction in a group of such ants. It is thus regarded as an example of
emergent phenomena~\cite{Vicsek1995} and the term `swarm
intelligence' has been coined to describe this idea. Various computational
techniques can be categorized as based on swarm intelligence (see, e.g.,
Refs.~\citeonline{Dorigo2004,Jafar2010}).
From a physical point of view, ants provide a good example of active
matter~\cite{Marchetti2013}, which can aggregate~\cite{Deneubourg2002} or
circulate~\cite{Couzin2003} spontaneously and exhibit peculiar mechanical
properties~\cite{Tennenbaum2015}.

The Patlak-Keller-Segel equation is a canonical starting point to study
organisms that interact by means of chemical
attractants~\cite{Patlak1953,Keller1970}. This model treats
the density of organisms $\rho(\mathbf{r},t)$ and the concentration of chemical
attractants $c(\mathbf{r},t)$ as continuous variables, where $\mathbf{r}$
denotes spatial coordinates and $t$ means time, and describes the interplay
between them. The Patlak-Keller-Segel equation has been extensively studied by
mathematicians and a variety of review papers are
available (see, e.g., Refs.~\citeonline{Horstmann2003,Horstmann2004}).
One of characteristic features of this model
is that the organisms can form a dense aggregate, developing a
$\delta$-function peak within a finite time, when the space has dimensionality
$d > 1$. Although such a `blow-up' phenomenon provides an approximate
description for biological aggregation, it is not entirely realistic that the
whole population collapses to a single point. Researchers have suggested various
mechanisms to regularize this singularity: To name a few, there are
density-dependent chemotactic
sensitivity~\cite{Biler1999,Hillen2001,Painter2002,Potapov2005}, nonlinear
diffusion~\cite{Kowalczyk2005,Kowalczyk2008},
logistic damping~\cite{Tello2007}, cross
diffusion~\cite{Hittmeir2011}, and shear flows~\cite{Bedrossian2016}. One may
also refer to a review by Hillen and Painter~\cite{Hillen2009} for many
variations of the classical Patlak-Keller-Segel model. One may also refer to
Ref.~\citeonline{Amorim2015} to see how it can be used to describe the
organization of a foraging path.

This work adopts the idea of nonlinear
diffusion~\cite{Kowalczyk2005,Kowalczyk2008} to take into account the finite
volume of the organisms, and analyse its consequences. Let us write down the
following set of equations:
\begin{eqnarray}
\frac{\partial \rho}{\partial t} &=& \nabla \cdot (- \chi_0 \rho \nabla c +
D_0 \rho \nabla \rho)\label{eq:kow1}\\
\frac{\partial c}{\partial t} &=& f_0 \rho + \nu_0 \nabla^2 c - g_0
c,\label{eq:kow2}
\end{eqnarray}
where $\chi_0$, $D_0$, $f_0$, $\nu_0$, and $g_0$ are positive constants.
The terms on the right-hand side of Eq.~(\ref{eq:kow1}) represent
chemotactic movement and nonlinear diffusion, respectively. On the other hand,
the three terms on the right-hand side of Eq.~(\ref{eq:kow2}) mean generation,
diffusion, and degradation, respectively.
According to the original derivation~\cite{Kowalczyk2005}, the nonlinear
diffusion term derives from $\rho \nabla h(\rho)$ with a pressure function
$h(\rho)$ due to crowding. If the pressure is expanded as a power series of
density, as in the virial expansion, the choice of $h(\rho) \propto \rho$
corresponds to the lowest-order approximation, because the zeroth order clearly
vanishes as $h(\rho = 0) = 0$. Some numerical observations have
been reported in this case~\cite{Kowalczyk2008,Hillen2009}.
Although $h(\rho)$ is effective pressure to describe collective motion
phenomenologically, it is interesting to note that an ant aggregate has an
elastic modulus, which has units of pressure, as a linear function of $\rho$,
until the ants are so densely packed that their legs are
compressed~\cite{Tennenbaum2015}.
Note that the classical Patlak-Keller-Segel equation is interpreted as $h(\rho)
\sim \ln \rho$ from this viewpoint.

In this work, we show that the system described by Eqs.~(\ref{eq:kow1}) and
(\ref{eq:kow2}) has a Lyapunov functional whose time derivative is smaller than
or equal to zero all the time. It will also be called the
free energy on the analogy with statistical mechanics. In general, a Lyapunov
functional is a powerful tool in analysing a dynamical system, and its existence
can be utilised to study properties of a fixed point beyond the local stability
analysis~\cite{Strogatz2001}.
After examining two stationary states, of
which one is homogeneous and the other is not, we investigate the Lyapunov
functional in the normal-mode coordinates to examine the transition between the
homogeneous and inhomogeneous states, restricting ourselves to radially
symmetric solutions.
We will minimize the Lyapunov functional with a Monte Carlo method
because it is computationally efficient in studying long-time behaviour of the
system. We then briefly check if the Monte Carlo results are consistent with
those from the direct numerical integration of the partial differential
equations. After characterizing the transition
based on the free-energy landscape, we conclude this work.

\section*{Analysis}

In this section, we begin with deriving the Lyapunov functional of
Eqs.~(\ref{eq:kow1}) and (\ref{eq:kow2}). We are interested in homogeneous and
inhomogeneous solutions and a transition between them. Of course, their
stability can be studied in a standard way by adding small perturbation with the
lowest nonzero mode, as will be demonstrated below. However, our main point is
that the transition from the homogeneous distribution to aggregation can be
analysed in detail by means of the Lyapunov functional, which contains the full
spectrum of possible modes in this system.

\subsection*{Lyapunov functional}

Before proceeding, we have to specify the boundary conditions of our model.
In analysing Eqs.~(\ref{eq:kow1}) and
(\ref{eq:kow2}), we consider a two-dimensional disc of radius $l$ and choose the
Neumann boundary conditions,
\begin{equation}
\frac{\partial \rho}{\partial r} = \frac{\partial c}{\partial r} = 0
\end{equation}
at $r = 0$ and $r=l$, where $r \equiv |\mathbf{r}|$ is the distance from the
origin of the disc. This condition means that the organisms cannot enter or
escape from the system across the boundary, which is the experimental situation
under consideration. In other words,
Eq.~(\ref{eq:kow1}) is derived from a continuity equation with current
$\mathbf{j} = - \chi_0 \rho \nabla c + D_0 \rho \nabla \rho$, which implies that
it conserves the total mass of the organisms:
\begin{equation}
M = \int_0^{2\pi} \int_0^l \rho (r, \theta) r~dr~d\theta
= \int \rho(r, \theta) dV,
\label{eq:mass}
\end{equation}
where $\theta$ means the angle in the polar coordinates and $dV$ is a volume
element.

If we assume that the chemical attractant reaches a stationary state very
quickly, so that the left-hand side of Eq.~(\ref{eq:kow2}) can be taken to be
approximately zero, we can solve the equation for $c$~\cite{Fatkullin2013}.
Let us consider the entire two-dimensional space for simplicity.
The formal solution is then given as
\begin{equation}
c(\mathbf{x}) = -\frac{f_0}{\nu_0} \int d\mathbf{y}
\mathcal{G}(\mathbf{x}-\mathbf{y}) \rho(\mathbf{y}),
\end{equation}
where $\mathcal{G}$ is the Green function obtained in terms of
$K_0$, the modified Bessel function of the second kind, as follows:
\begin{equation}
\mathcal{G}(\mathbf{x} - \mathbf{y}) = -\frac{1}{2\pi} K_0 \left( \kappa |
\mathbf{x} - \mathbf{y} | \right)
\label{eq:green}
\end{equation}
with $\kappa \equiv \sqrt{f_0 / \nu_0}$. Plugging this into Eq.~(\ref{eq:kow1}),
we find that
\begin{equation}
\frac{\partial \rho}{\partial t} = \nabla \cdot \left( \rho \nabla \frac{\delta
\mathcal{E}}{\delta \rho} \right)
\label{eq:relax}
\end{equation}
with
\begin{equation}
\mathcal{E} \equiv \frac{D_0}{2} \int \rho^2 (\mathbf{x}) d\mathbf{x} +
\frac{f_0 \chi_0}{2\nu_0} \iint \rho(\mathbf{x}) \mathcal{G}(\mathbf{x} -
\mathbf{y}) \rho(\mathbf{y}) d\mathbf{x} d\mathbf{y}.
\label{eq:energy}
\end{equation}
Note that the first term is equivalent to the participation ratio in the
localization problem~\cite{Edwards1972}, and the second term can be interpreted
as interaction energy between organisms at a distance. The participation ratio
is minimized when $\rho$ is distributed homogeneously, whereas the effective
interaction potential, Eq.~(\ref{eq:green}), make the organisms attract
each other. If diffusion is dominant, i.e., $D_0 \nu_0 \gg f_0 \chi_0$, the
interaction term becomes negligible and the aggregation mediated by the
chemical attractants will be suppressed.
From Eqs.~(\ref{eq:relax}) and (\ref{eq:energy}), it is
straightforward to see that
\begin{equation}
\frac{d\mathcal{E}}{dt} = - \int \left| \nabla \frac{\delta \mathcal{E}}{\delta
\rho} \right|^2 \rho(\mathbf{x}) d\mathbf{x},
\end{equation}
which implies that $\mathcal{E}$ never increases as time goes by.

We have derived Eq.~(\ref{eq:energy}) under the restriction that $\partial
c / \partial t = 0$ only because $\mathcal{E}$ provides a simple physical
interpretation in terms of $\rho$ only. In fact, it is possible to construct a
complete Lyapunov functional without such a restriction:
Let us rescale the variables as $\tau = D_0 t$ and $c' = \frac{\chi_0}{D_0} c$.
to rewrite Eqs.~(\ref{eq:kow1}) and (\ref{eq:kow2}) as
\begin{eqnarray}
\frac{\partial \rho}{\partial \tau} &=& \nabla \cdot (-\rho \nabla c' +
\rho \nabla \rho) = \nabla \cdot (\rho \nabla Z)\\
\frac{\chi_0}{\nu_0} \frac{\partial c'}{\partial \tau} &=&
\nabla^2 c' - \frac{g_0}{\nu_0} c' + \frac{f_0 \chi_0}{D_0 \nu_0} \rho,
\end{eqnarray}
where $Z \equiv \rho - c'$. We can show that
\begin{equation}
\int dV~ Z \frac{\partial \rho}{\partial \tau} = \int dV~ \nabla \cdot (Z
\rho \nabla Z) -\int dV~ \rho |\nabla Z|^2,
\label{eq:Z}
\end{equation}
where the first term on the right-hand side vanishes due to the boundary
conditions. By using Eq.~(\ref{eq:Z}), we can also show the following:
\begin{eqnarray}
\frac{d}{d\tau} \int dV~\rho Z &=& \int dV~ \rho \frac{\partial Z}{\partial
\tau} + \int dV~ Z\frac{\partial \rho}{\partial \tau}\\
&=& \frac{d}{d\tau} \int dV~ \frac{\rho^2}{2} - \int dV~ \rho \frac{\partial
c'}{\partial \tau} - \int dV~ \rho |\nabla Z|^2.
\label{eq:rhoZ}
\end{eqnarray}
In addition, we have the following equality:
\begin{eqnarray}
0 &=& \int dV~ \nabla \cdot \left( \frac{\partial c'}{\partial \tau} \nabla c'
\right)\\
&=& \frac{\chi_0}{\nu_0} \int dV~ \left( \frac{\partial
c'}{\partial \tau} \right)^2
+ \frac{g_0}{\nu_0} \int dV~ c'\frac{\partial c'}{\partial \tau}
- \frac{f_0 \chi_0}{D_0 \nu_0} \int dV~ \rho \frac{\partial c'}{\partial
\tau}
+ \frac{d}{d\tau} \int dV~ \frac{\left| \nabla c' \right|^2}{2}.
\label{eq:lya}
\end{eqnarray}
Plugging Eq.~(\ref{eq:rhoZ}) into Eq.~(\ref{eq:lya}), we get
\begin{equation}
-\frac{dW}{d\tau} = \frac{\chi_0}{\nu_0} \int dV \left( \frac{\partial
c'}{\partial \tau} \right)^2 + \frac{f_0 \chi_0}{D_0 \nu_0} \int dV~ \rho
|\nabla Z|^2,
\label{eq:dW}
\end{equation}
where
\begin{equation}
W \equiv \frac{f_0 \chi_0}{D_0 \nu_0} \int dV~ \left( \frac{1}{2} \rho^2 -
\rho c' \right) + \frac{g_0}{2\nu_0} \int dV~ |c'|^2 + \int dV~
\frac{|\nabla c'|^2}{2}.
\label{eq:W}
\end{equation}
It is clear from Eq.~(\ref{eq:dW}) that $dW/d\tau$ cannot be positive so that
$W$ does not increase when the system evolves according to Eqs.~(\ref{eq:kow1})
and (\ref{eq:kow2}). For this reason, this quantity is sometimes called the
\emph{free energy} of this system. The time derivative $dW/d\tau$ equals zero if
$\partial c'/\partial \tau=0$ and $\mathbf{j} \propto \nabla Z =0$ everywhere
that $\rho>0$.
The first integral of Eq.~(\ref{eq:W}) consists of the
participation ratio and the potential energy due to the
coupling between $\rho$ and $c$, whereas the other two integrals describe the
chemical energy~\cite{Calvez2008}.
Likewise, one can argue that Eq.~(\ref{eq:dW}) contains the chemical production
term $\propto (\partial c/\partial t)^2$ on its right-hand side, and that the
last term corresponds to something referred to as entropy production in the
classical Patlak-Keller-Segel model because it is related to
the time derivative of the Shannon entropy~\cite{Calvez2008}.
In our nonlinear-diffusion model, the last term of Eq.~(\ref{eq:dW}) may be
regarded as generalized entropy production in terms of the Tsallis
entropy~\cite{Tsallis1988}.
It is also worth noting that the integrands in Eq.~(\ref{eq:W})
are all quadratic, which will turn out to be useful for our analysis.

\subsection*{Linear stability of a homogeneous stationary solution}
Equations~(\ref{eq:kow1}) and (\ref{eq:kow2}) admit a homogeneous stationary
solution $\rho = \frac{g_0}{f_0} c = \rho_{\rm const}$, where $\rho_{\rm const}
= M/(\pi l^2)$ from Eq.~(\ref{eq:mass}). In this state, Eq.~(\ref{eq:W}) yields
\begin{equation}
W = \frac{M^2}{2 \pi l^2} \frac{f_0 \chi_0}{D_0 \nu_0} \left( 1 -
\frac{f_0 \chi_0}{D_0 g_0} \right).
\label{eq:lhom}
\end{equation}
The standard linear stability analysis
assumes small perturbations $\epsilon_\rho$ and $\epsilon_c$ around this
homogeneous solution to assume $\rho(\mathbf{r},t) = \rho_{\rm const} +
\epsilon_\rho (\mathbf{r},t)$ and $c(\mathbf{r},t) = \frac{f_0}{g_0} \rho_{\rm
const}
+ \epsilon_c (\mathbf{r},t)$. By collecting linear terms in $\epsilon_\rho$ and
$\epsilon_c$, we obtain
\begin{equation}
\frac{\partial}{\partial t}
\begin{pmatrix}
\epsilon_\rho \\ \epsilon_c
\end{pmatrix}
=
\begin{pmatrix}
0 & 0\\
f_0 & -g_0
\end{pmatrix}
\begin{pmatrix}
\epsilon_\rho \\ \epsilon_c
\end{pmatrix}
+
\begin{pmatrix}
D_0 \rho_{\rm const} & -\chi_0 \rho_{\rm const}\\
0 & \nu_0
\end{pmatrix}
\nabla^2
\begin{pmatrix}
\epsilon_\rho \\ \epsilon_c
\end{pmatrix}.
\end{equation}
Suppose that the perturbations are described as cylindrical harmonics,
satisfying the following equation:
\begin{equation}
(\nabla^2 + k^2)
\begin{pmatrix}
\epsilon_\rho \\ \epsilon_c
\end{pmatrix} = 0.
\end{equation}
Each mode then takes the form of $J_n (kr) e^{\pm in\theta} e^{\eta t}$, where
$J_n$ means the Bessel function and $\eta$ is its growth rate. The Neumann
boundary conditions are expressed as $\frac{\partial}{\partial r} J_n(kl) = 0$.
The lowest mode is thus found at $n=0$, which means radially symmetric density
fluctuations concentrated around the origin. The first zero of $J_1$ is located
at $kl \approx 3.832\ldots$. If we solve the resulting eigenvalue problem:
\begin{equation}
\eta^2 + [k^2(D_0 \rho_{\rm const} + \nu_0) + g_0] \eta +
\rho_{\rm const} [k^2 D_0 (g_0 + k^2\nu_0) - k^2f_0\chi_0] = 0,
\end{equation}
the stability condition is obtained as
$k^2 D_0 (g_0 + k^2\nu_0) - k^2f_0\chi_0 > 0$.
Note that it is independent of $\rho_{\rm const}$, differently from the
classical Patlak-Keller-Segel model~\cite{Childress1981},
so that the system does
not need critical mass for instability. This feature is, however, due to our
particular choice of nonlinear diffusion.
We find a necessary condition for the lowest mode to grow in time as follows:
\begin{equation}
k^2 l^2 \approx 14.684 < \left( \frac{f_0 \chi_0}{D_0 \nu_0} - \frac{g_0}{\nu_0}
\right) l^2 = K^2 l^2,
\label{eq:instability}
\end{equation}
where
\begin{equation}
K \equiv \sqrt{\frac{f_0 \chi_0}{D_0 \nu_0} - \frac{g_0}{\nu_0}}.
\label{eq:K}
\end{equation}
If we assume that $g_0 \ll 1$, the expression inside the square root of
Eq.~(\ref{eq:K}) is interpreted as a ratio between chemotactic strength and
diffusivity.
This small-$g_0$ limit is often plausible without altering the essential
physics, because some ant pheromones last for days~\cite{Regnier1968}.
Equation~(\ref{eq:instability}) suggests that $Kl$ will be an
important dimensionless parameter that governs the aggregation phenomenon.

In addition,
if the disc is so large that the boundary effects are negligible and there is
a continuous spectrum of possible wavenumbers,
the initial stage of instability from the homogeneous solution is governed by
the most unstable mode with $k=k_u$ such that maximizes the positive
$\eta$~\cite{Potapov2005}.
The wavenumber $k_u$ can be expressed by the following formula:
\begin{equation}
k_u^2 = f_0 \chi_0 \sqrt{\frac{\rho_{\rm const}}{D_0 \nu_0}} \left(
\frac{1}{\sqrt{D_0 \rho_{\rm const}} + \sqrt{\nu_0}} \right)^2,
\label{eq:ku}
\end{equation}
where we take the limit of $g_0 \rightarrow 0$ to simplify the expression.
Equation~(\ref{eq:ku}) will determine the typical length scale between
aggregates, when the homogeneous initial state becomes unstable.

\subsection*{Inhomogeneous stationary solution}
Let us now consider a radially symmetric stationary aggregate. The boundary
conditions make the flux vanish everywhere, i.e., $\mathbf{j} = -\chi_0 \rho
\nabla c + D_0 \rho \nabla \rho = 0$. It implies that
\begin{equation}
\rho = \frac{\chi_0}{D_0} (c - c_0)
\label{eq:rho_c}
\end{equation}
with a constant of integration $c_0$.
Substituting Eq.~(\ref{eq:rho_c}) into Eq.~(\ref{eq:kow2}) with the stationarity
condition, we obtain an inhomogeneous Helmholtz equation:
\begin{equation}
0 = \frac{f_0 \chi_0}{D_0} (c-c_0) + \nu_0 \nabla^2 c - g_0 c,
\label{eq:helmholtz}
\end{equation}
which has the following radially symmetric solution:
\begin{equation}
c(r) = A J_0 (Kr) + \frac{f_0 \chi_0}{K^2 D_0 \nu_0} c_0,
\label{eq:c}
\end{equation}
where $A$ is a constant describing the amplitude of aggregation,
$J_n$ is the Bessel function, and the wavenumber $K$ has been defined in
Eq.~(\ref{eq:K}) above.
Obviously, the solution is feasible only when the boundary condition
is satisfied by $\left. \frac{d}{dr} J_0 (Kr) \right|_{r=l} = -KJ_1 (Kl) = 0$,
and let us suppose that this is the case.
The constant $A$ is bounded by a condition that both $\rho$ and $c$ must be
non-negative everywhere.
If we plug Eq.~(\ref{eq:c}) into Eq.~(\ref{eq:rho_c}), we find that
\begin{equation}
\rho(r) = \frac{\chi_0}{D_0} \left[ AJ_0 (Kr) + \left( \frac{f_0 \chi_0}{K^2 D_0
\nu_0} - 1 \right) c_0 \right].
\end{equation}
The unknown constant $c_0$ can be explicitly determined from Eq.~(\ref{eq:mass})
because $\int_0^l J_0 (Kr) r~dr = 0$ as long as the boundary conditions are
satisfied. After some algebra, we can write the results as
\begin{eqnarray}
\rho &=& \frac{\chi_0}{D_0} A J_0 + \rho_{\rm const}\\
c &=& A J_0 + c_{\rm const},
\end{eqnarray}
where $\rho_{\rm const}$ and $c_{\rm const}$ define the homogeneous solution.
We substitute these results into Eq.~(\ref{eq:W}) to calculate the Lyapunov
functional:
\begin{eqnarray}
W &=& \frac{f_0 \chi_0}{D_0 \nu_0} \int dV \left[
\frac{1}{2} \left( \frac{\chi_0}{D_0}
AJ_0 + \rho_{\rm const}\right)^2 - \left( \frac{\chi_0}{D_0} AJ_0 + \rho_{\rm
const} \right)
\frac{\chi_0}{D_0} \left( AJ_0 + c_{\rm const} \right)
\right]\nonumber\\
&& + \frac{g_0}{2\nu_0} \int dV \left[ \frac{\chi_0^2}{D_0^2} \left( AJ_0 +
c_{\rm const} \right)^2 \right]
+ \frac{1}{2} \int dV \frac{\chi_0^2}{D_0^2} (KAJ_1)^2\\
&=& -\frac{f_0 \chi_0^3 A^2}{2 D_0^3 \nu_0} \int dV J_0^2 + \left( \frac{1}{2}
\frac{f_0 \chi_0}{D_0 \nu_0} \rho_{\rm const}^2 - \frac{f_0 \chi_0^2}{D_0^2
\nu_0} \rho_{\rm const} c_{\rm const} \right) \pi l^2 \nonumber\\
&& + \frac{g_0 \chi_0^2 A^2}{2 D_0^2 \nu_0} \int dV J_0^2 + \frac{\chi_0^2
g_0}{2 D_0^2 \nu_0} c_{\rm const}^2 \pi l^2 + \frac{\chi_0^2 K^2 A^2}{2 D_0^2}
\int dV J_1^2.
\label{eq:W_twomode}
\end{eqnarray}
We can see that the three integrals on the last line vanish altogether, if we
note the definition of $K$ [Eq.~(\ref{eq:K})] and the following identity:
\begin{equation}
K^2 \int_0^l r J_0^2 (Kr) dr = K \int_0^l \frac{d}{dr} [r J_1 (Kr)] J_0 (Kr) dr
= - K \int_0^l rJ_1 (Kr) \frac{d}{dr} J_0(Kr) dr = K^2 \int_0^l r J_1(Kr)^2 dr,
\label{eq:j0j1}
\end{equation}
which is valid under our assumption that $J_1(Kl) = 0$.
As a result, we obtain
\begin{equation}
W = \frac{M^2}{2 \pi l^2} \frac{f_0 \chi_0}{D_0 \nu_0} \left( 1-
\frac{f_0 \chi_0}{g_0 D_0} \right),
\end{equation}
which is identical to the Lyapunov functional of the homogeneous solution
[Eq.~(\ref{eq:lhom})]. It is consistent with the fact that the solution with $K$
has neutral stability in the linear-stability analysis [see, e.g.,
Eq.~(\ref{eq:instability})], according to which the radially symmetric mode
$\propto J_0(kr)$ can survive only when $k$ is smaller than $K$.
Although we have assumed that the wavenumber $K$ is compatible with the boundary
condition, it is actually independent of $l$, which implies that the
stationarity condition cannot be met exactly.
If a perturbative mode with $k < K$ appears from the homogeneous state with
satisfying the boundary conditions, therefore,
it cannot be stationary: Its amplitude will grow exponentially at
first, but cannot become arbitrarily large because of the non-negativity of
$\rho$ and $c$. The growth will stop when $A$ reaches the largest value that
does not violate the non-negativity. This scenario seems to suggest a
jump in $A$ as $K$ crosses a threshold, and this scenario will be scrutinized
below by considering a full spectrum of normal modes.

\subsection*{Normal-mode expansion}

Let us decompose $\rho$ and $c$ into normal modes:
\begin{eqnarray}
\rho(r,\theta,t) &=& \rho_{\rm const} + \sum_{p=0}^\infty \sum_{m=1}^\infty
J_p(j'_{p,m}r/l) \left[ E_{pm}(t) \cos p\theta + F_{pm}(t) \sin p\theta \right]
\label{eq:rho_basis}\\
c(r,\theta,t) &=& c_{\rm const} + \sum_{p=0}^\infty \sum_{m=1}^\infty
J_p(j'_{p,m}r/l) \left[ G_{pm}(t) \cos p\theta + H_{pm}(t) \sin p\theta \right],
\label{eq:c_basis}
\end{eqnarray}
where $j'_{pm}$ denotes the $m$th zero of $\frac{d}{dx} J_p(x)$.
Note that
\begin{equation}
\int_0^{2\pi} \int_0^l r J_p(j'_{pm} r/l) e^{i p\theta} ~dr ~d\theta = 0,
\end{equation}
so that Eq.~(\ref{eq:rho_basis}) automatically conserves the total mass
$M = \int_0^{2\pi} \int_0^l \rho(r,\theta) r~dr~d\theta = \rho_{\rm const} \pi
l^2$. Likewise, the total amount of the chemical attractant is given as $c_{\rm
const} \pi l^2$, which is, however, a function of time in general.
It is straightforward to see the following orthogonality relation
\begin{equation}
\int_0^l r J_p(j'_{pu}r/l) J_p(j'_{pw}r/l) ~dr
= -\frac{l^2}{2} J_p (j'_{pu}) \frac{d^2}{dx^2} J_p (j'_{pu}) \delta_{uw}
= -\frac{l^2}{2} \phi_{pu} \delta_{uw},
\label{eq:ortho}
\end{equation}
where $\delta_{uw}$ is the Kronecker delta and $\phi_{pu} \equiv
J_p (j'_{pu}) \frac{d^2}{dx^2} J_p (j'_{pu})$.

We will rewrite the Lyapunov functional [Eq.~(\ref{eq:W})] by using
Eqs.~(\ref{eq:rho_basis}) and (\ref{eq:c_basis})]. The first term needs
an integral of $\rho^2$ over the disc, which can be expressed as
\begin{equation}
\frac{1}{\pi l^2}
\int \rho^2 dV = \rho_{\rm const}^2 - \sum_{m=1}^\infty \left[ \phi_{0m}
E_{0m}^2 + \frac{1}{2} \sum_{p=1}^\infty \phi_{pm} \left( E_{pm}^2 + F_{pm}^2
\right) \right]
\label{eq:rho2}
\end{equation}
by using the orthogonality relations. The integrals of $\rho c$ and $c^2$ can be
done in a similar way. However, the last part of the Lyapunov functional
[Eq.~(\ref{eq:W})] is more complicated: It is involved with an integral of
$|\nabla c|^2$, which is decomposed into two terms:
\begin{equation}
\int |\nabla c|^2 dV =
\int_0^{2\pi} \int_0^l
\left| \frac{\partial c}{\partial r} \right|^2 r~dr~ d\theta
+ \int_0^{2\pi} \int_0^l
\frac{1}{r^2} \left| \frac{\partial c}{\partial \theta} \right|^2 r~dr~ d\theta.
\end{equation}
We again substitute Eqs.~(\ref{eq:rho_basis}) and (\ref{eq:c_basis}) here to
obtain
\begin{eqnarray}
\int_0^{2\pi} \int_0^l
\left| \frac{\partial c}{\partial r} \right|^2 r~dr~ d\theta
&=&
\pi \sum_{m=1}^\infty j_{1m}^2 J_0^2(j_{1m}) G_{0m}^2\nonumber\\
&+& \frac{\pi}{l^2} \sum_{p=1}^\infty \sum_{mn}
j'_{pm} j'_{pn} (G_{pm} G_{pn} + H_{pm} H_{pn})
\int_0^l r \left( \left. \frac{dJ_p(x)}{dx} \right|_{x=j'_{pm}r/l} \right)
\left( \left. \frac{dJ_p(x)}{dx} \right|_{x=j'_{pn}r/l} \right) dr
\end{eqnarray}
and
\begin{equation}
\int_0^{2\pi} \int_0^l
\frac{1}{r^2} \left| \frac{\partial c}{\partial \theta} \right|^2 r~dr~ d\theta
= \pi \sum_{p=1}^\infty \sum_{mn} p^2 (G_{pm} G_{pn} + H_{pm} H_{pn}) \int_0^l
\frac{1}{r} J_p (j'_{pm}r/l) J_p (j'_{pn}r/l) dr.
\end{equation}
Note that the results still have the triple sums over $p$, $m$, and $n$, because
we cannot enjoy the orthogonality between $m$ and $n$ when performing the
integrals over $r$.

\begin{figure}
\includegraphics[width=\textwidth]{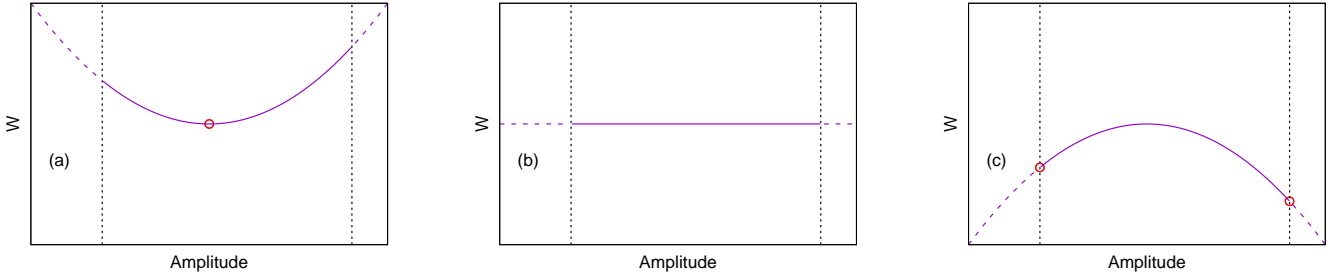}
\caption{Sketches of the Lyapunov functional $W$ along a principal axis, a
combination of the amplitudes $E_{01}$ and $G_{01}$, when (a) $Kl < j_{11}$, (b)
$Kl = j_{11}$, and (c) $Kl > j_{11}$, respectively.
The vertical dotted lines represent the physical
constraint that both $\rho$ and $c$ should be non-negative, so that the system
can explore only the landscapes of $W$ drawn with solid lines. The small red
circles show local minima of the given landscapes.}
\label{fig:scheme}
\end{figure}

To circumvent the time-consuming evaluation of the triple sums,
we focus on radially symmetric solutions by setting $p=0$.
If $j_{pm}$ denotes the $m$th zero of $J_p(x)$, we can identify $j'_{0m}$ with
$j_{1m}$ because $\frac{d}{dx} J_0(x) = -J_1(x)$. Therefore,
Eq.~(\ref{eq:ortho}) further simplifies to
\begin{equation}
\int_0^l r J_0(j_{1u}r/l) J_0(j_{1w}r/l) ~dr
= \int_0^l r J_1(j_{1u}r/l) J_1(j_{1w}r/l) ~dr
= \frac{l^2}{2} J_0^2 (j_{1u}) \delta_{uw},
\label{eq:ortho2}
\end{equation}
where the first equality is derived in the same way as in Eq.~(\ref{eq:j0j1}),
and the second one is the conventional orthogonality of the Bessel
function~\cite{Boas2006}.
Plugging Eqs.~(\ref{eq:rho_basis}) and (\ref{eq:c_basis}) with $p=0$
into the Lyapunov functional [Eq.~(\ref{eq:W})] and using the orthogonality,
we find that
\begin{eqnarray}
\frac{W}{\pi l^2} &=& \frac{f_0 \chi_0}{D_0 \nu_0} \left[ \frac{1}{2} \left(
\rho^2_{\rm const} + \sum_{m=1}^\infty J_0^2(j_{1m}) E_{0m}^2 \right)
-\frac{\chi_0}{D_0} \left( \rho_{\rm const} c_{\rm const} + \sum_{m=1}^\infty
J_0^2(j_{1m}) E_{0m} G_{0m} \right) \right]\nonumber\\
&& + \frac{g_0 \chi_0^2}{2 D_0^2 \nu_0} \left( c_{\rm const}^2 +
\sum_{m=1}^\infty J_0^2(j_{1m}) G_{0m}^2 \right) + \frac{\chi_0^2}{2 D_0^2 l^2}
\sum_{m=1}^\infty j_{1m}^2 J_0^2(j_{1m}) G_{0m}^2\\
&=& \sum_{m=0}^\infty
\frac{1}{2} J_0^2(j_{1m})
\begin{pmatrix}
E_{0m} & \frac{\chi_0}{D_0} G_{0m}
\end{pmatrix}
\begin{pmatrix}
\frac{f_0 \chi_0}{D_0 \nu_0} & - \frac{f_0 \chi_0}{D_0 \nu_0} \\
- \frac{f_0 \chi_0}{D_0 \nu_0} & \frac{g_0}{\nu_0} + \frac{j_{1m}^2}{l^2}\\
\end{pmatrix}
\begin{pmatrix}
E_{0m} \\ \frac{\chi_0}{D_0} G_{0m}
\end{pmatrix},
\label{eq:Wp0}
\end{eqnarray}
where we have defined $E_{00} \equiv \rho_{\rm const}$, $G_{00} \equiv c_{\rm
const}$, and $j_{10} \equiv 0$.
We are interested in the minimum of Eq.~(\ref{eq:Wp0}), expecting that it
captures the long-term behaviour of the system.
The set of variables $\{ c_{\rm const}, E_{01},
E_{02}, \ldots, G_{01}, G_{02}, \ldots \}$ resulting from the
minimization will be independent of the overall rescaling of $W$ and thus
determined by three dimensionless ratios, $\chi_0/D_0$, $g_0/f_0$, and
$\nu_0/(f_0 l^2)$. The first ratio measures the chemical sensitivity of the
organism with respect to its nonlinear diffusivity. The next one measures the
relative time scale between the generation and decay of the chemical attractant.
Finally, the last one gives the typical time scale for the chemical attractant
to diffuse into the whole system, measured with respect to the generation rate.
Let us assume that each summand
can be considered \emph{separately}
in this minimization problem. Then, for $m=0$,
only $c_{\rm const}$ varies, because $\rho_{\rm const}$ is fixed by the total
mass $M$, and the optimal value for $c_{\rm const}$ equals $(f_0 / g_0)
\rho_{\rm const}$ as we have already seen in the homogeneous stationary
solution. For every other $m > 1$, we have a simple quadratic function
of $E_{0m}$ and $G_{0m}$. From an eigenvalue analysis, it is straightforward to
see that the functional shape is elliptic when $j_{1m} > Kl$ and hyperbolic
otherwise, where $K$ is defined by Eq.~(\ref{eq:K}).
In the former case, the minimum is located at $E_{0m} = G_{0m} = 0$.
In the latter case, the minima of Eq.~(\ref{eq:Wp0}) are found at $E_{0m}
\propto G_{0m} = \pm \infty$, and the divergence must be regulated by the
condition that both $\rho$ and $c$ are non-negative everywhere.
The idea is sketched in Fig.~\ref{fig:scheme} for $m=1$.
According to this argument,
if $Kl$ lies between $j_{11}$ and $j_{12}$, for example, we will observe two
local minima, one for $E_{01} \propto G_{01} > 0$ and the other for $E_{01}
\propto G_{01} < 0$, while all the other $E_{0m}$'s and $G_{0m}$'s with $m>1$
remain suppressed to zero.
An interesting point in this picture is that the Lyapunov functional becomes
independent of the amplitude of aggregation if $Kl$ exactly equals $j_{11}$:
An infinite number of states would have the same value of the Lyapunov
functional. Therefore, even if the system converges to two different states
as $Kl \rightarrow j_{11}^+$ and $Kl \rightarrow j_{11}^-$, respectively,
there would be a continuous spectrum of states between them at $Kl = j_{11}$.

\section*{Numerical results}

\begin{figure}
{\centering
\includegraphics[width=0.8\textwidth]{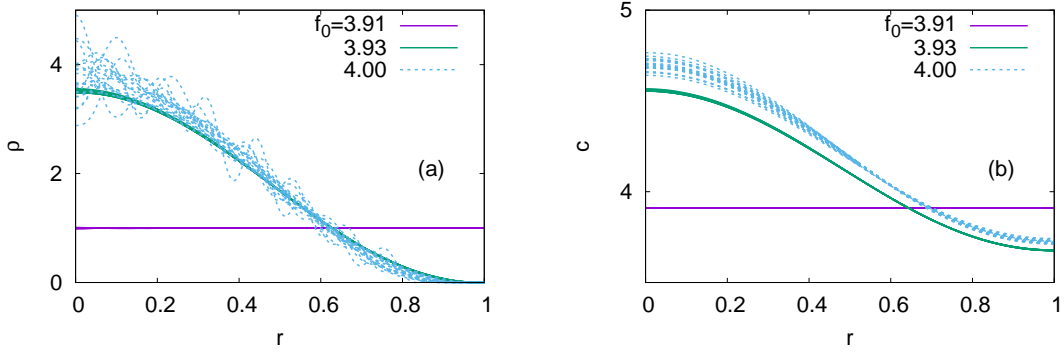}
\par}
\caption{(a) Density of the organisms $\rho$ and (b) the density of their
chemical attractants $c$, obtained by minimizing a partial sum of
Eq.~(\ref{eq:Wp0}) up to $m=19$
with the Metropolis algorithm. We choose $\chi_0 = 4$, $\rho_{\rm const} = 1$,
$D_0 = 1$, $\nu_0 = 1$, $g_0 = 1$, and $l=1$.
For each $f_0$, we run $20$ independent samples,
slowly lowering the `temperature' from $T=10$ to $T =0$.}
\label{fig:threshold}
\end{figure}

\begin{figure}
{\centering
\includegraphics[width=0.65\textwidth,angle=270]{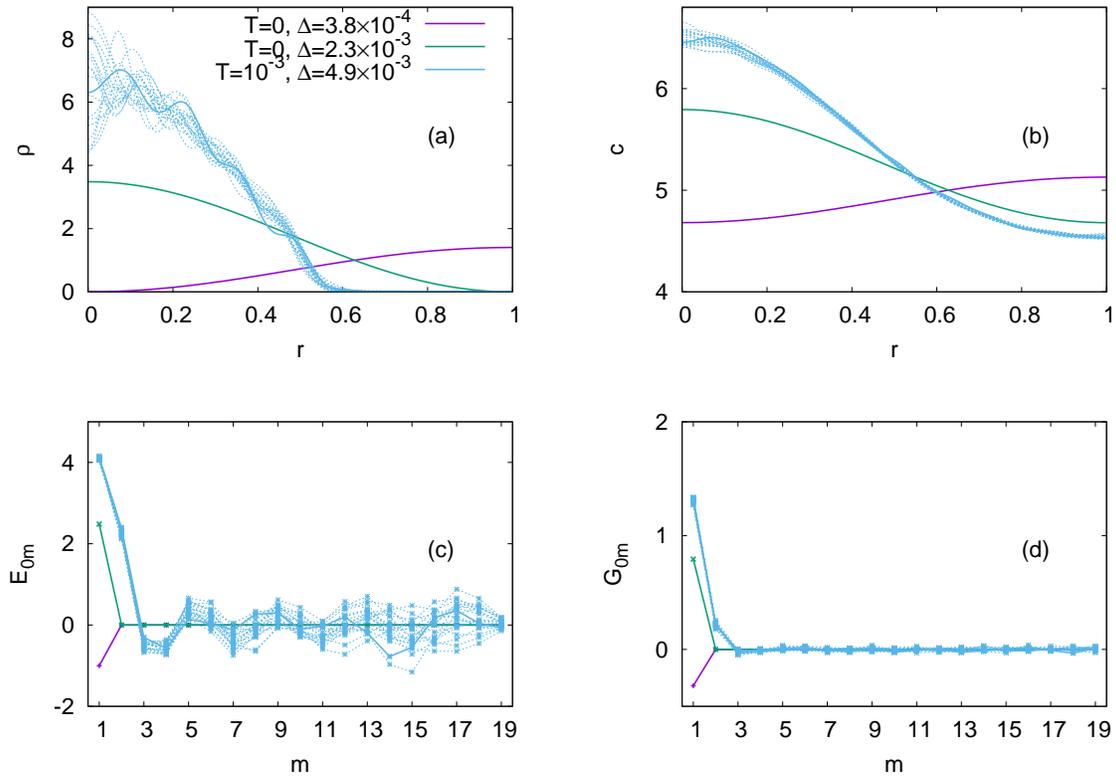}
\par}
\caption{Minimization results of a partial sum of Eq.~(\ref{eq:Wp0}) up to
$m=19$, obtained by the Metropolis algorithm. In this plot, we show (a) the
density of the organisms $\rho$, (b) that of the chemical attractants
$c$, (c) the amplitudes $E_{0m}$'s for describing $\rho$,
and (d) $G_{0m}$'s for $c$. We choose $f_0 = 5$, $\chi_0
= 4$, $\rho_{\rm const} = 1$, $D_0 = 1$, $\nu_0 = 1$, $g_0 = 1$, and
$l=1$. The initial condition is given by $c_{\rm const} = (f_0/g_0) \rho_{\rm
const}$ and $E_{0m} = G_{0m} = 0$ in each case. For the zero-temperature case,
i.e., $T=0$, the system approaches either of two different local minima,
represented by the purple and green lines, respectively. If we instead
slowly cool down the system from $T=10^1$ to $T \approx 10^{-3}$, we find high
concentrations of $\rho$ and $c$ around $r=0$ for all the $20$ samples shown in
this plot (the blue lines). Among the blue lines, the solid ones represent the
sample with the best minimization result.}
\label{fig:N20f5}
\end{figure}

\begin{figure}
{\centering
\includegraphics[width=0.65\textwidth, angle=270]{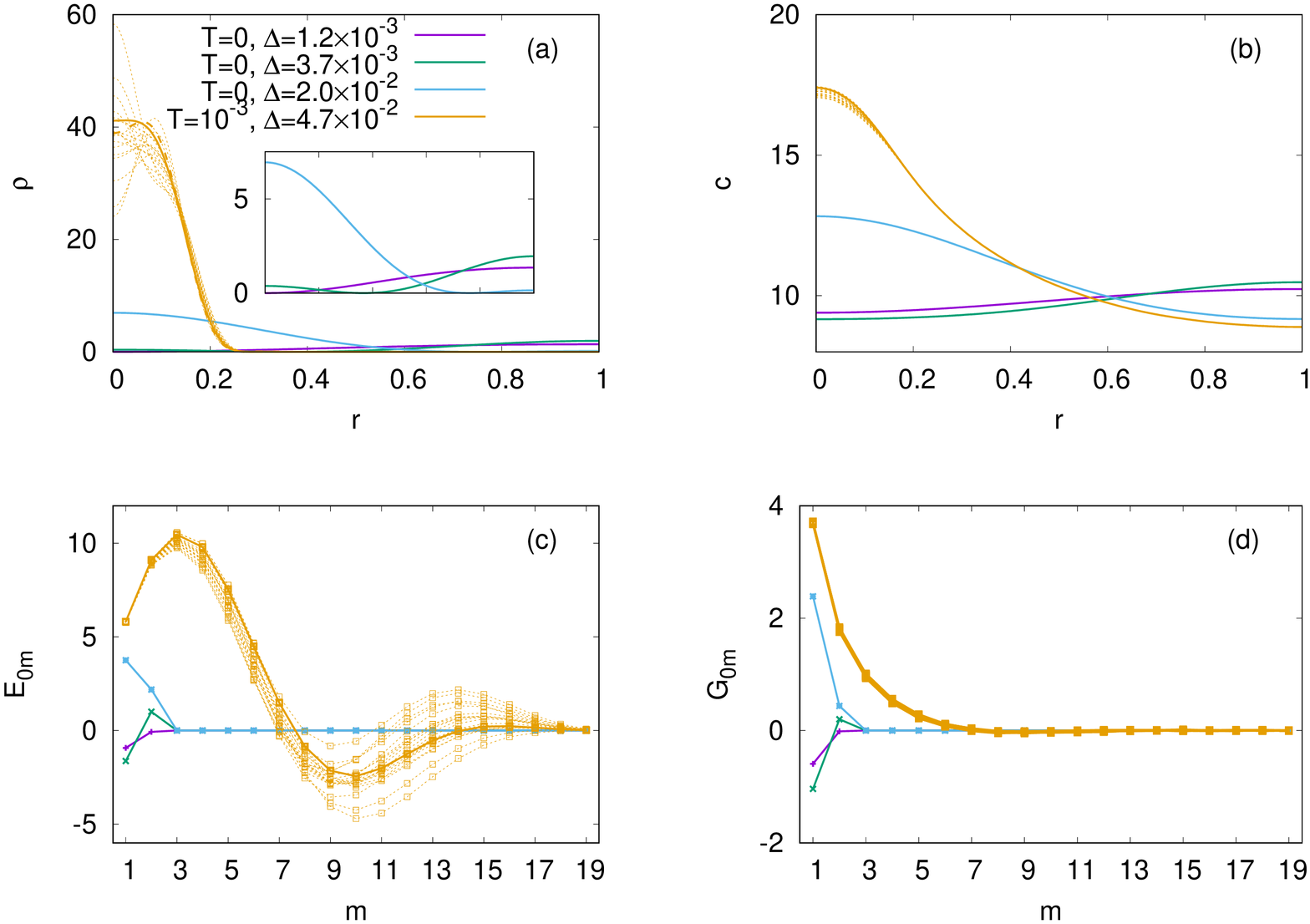}
\par}
\caption{Minimization results of a partial sum of Eq.~(\ref{eq:Wp0}) up to
$m=19$, obtained by the Metropolis algorithm. We choose $f_0 = 10$ and
$\chi_0 = 8$, and keep all the others the same as in Fig.~\ref{fig:N20f5}.
(a) The density of the organisms. (Inset)
If we run the zero-temperature Metropolis algorithm starting with $E_{01} =
E_{02} = \ldots = G_{01} = G_{02} = \ldots = 0$,
the system approaches either of three different local minima, which are
represented by the purple, green, and blue lines, respectively.
We can also start from $T=10$ and then cool down the system slowly.
Performing this process with $20$ independent samples, we plot their $\rho$ at
$T=10^{-3}$ with the orange lines. Among the orange lines, the solid ones
represent the sample with the best minimization result. The other panels show
(b) the density of the chemical attractants, (c) the normal-mode amplitudes for
$\rho$, and (d) those for $c$, respectively.}
\label{fig:N20f10}
\end{figure}

Let us choose $\chi_0=4$ and set other parameters, $\rho_{\rm const}$, $D_0$,
$\nu_0$, $g_0$, and $l$, to unity. With these parameters, the system reaches the
threshold for aggregation, $Kl = j_{11}$, when $f_0=f_0^\ast \approx 3.92$. We
minimize the Lyapunov function for radially symmetric cases [Eq.~(\ref{eq:Wp0})]
with different values of $f_0$ by means of the Metropolis algorithm (see Method
for details).
In evaluating Eq.~(\ref{eq:Wp0}) numerically, we have to replace the infinite
series by a partial sum, and the spatial resolution of the resulting expression
will be enhanced as we include more and more modes in the summation.
Here, let us use a partial sum up to $m=19$ because it already captures
the overall behaviour correctly. This choice implies that we have to work with
$39$ variables of $c_{\rm const}, E_{01}, \ldots, G_{0m}$.
For the algorithm to search for the parameter space efficiently, we
introduce a `temperature' variable $T$,
which helps the system escape from metastable local minima.
We start with a sufficiently high temperature, say, $T = 10^1$, to
explore a wide region of the parameter space and then gradually lower the
temperature down to $T=0$. As argued above, we observe a sharp transition from a
homogeneous solution to aggregation when $f_0$ exceeds $f_0^\ast \approx 3.92$,
and the aggregation pattern is approximated to $J_0 (j_{11}r/l)$
[Fig.~\ref{fig:threshold}].
From $f_0 = 3.93$ to $f_0=4.00$, on the other hand,
the system remains qualitatively the same, although small variations exist
from sample to sample.
To sum up, the behaviour at $Kl \approx j_{11}$
is indeed explained by the assumption that the minimization of
Eq.~(\ref{eq:Wp0}) can be carried out term by term.

\begin{figure}
{\centering
\includegraphics[width=0.45\textwidth]{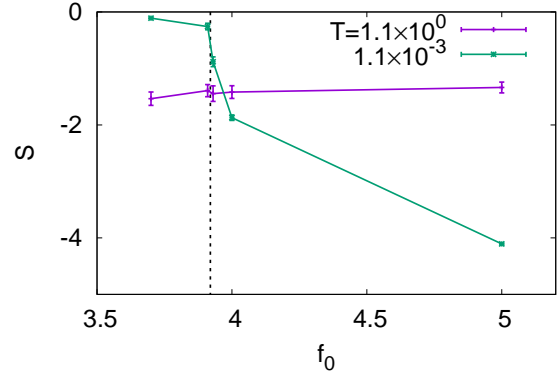}
\par}
\caption{Shannon entropy [Eq.~(\ref{eq:entropy})] as a function of $f_0$. The
other parameters are the same as in Figs.~\ref{fig:threshold} and
\ref{fig:N20f5}. For each data point, we take an average over $20$ independent
samples. The vertical dotted line represents $f_0 = f_0^\ast \approx 3.92$ to
make $Kl = j_{11}$.}
\label{fig:entropy}
\end{figure}

As $f_0$ increases, however, the assumption loses validity.
In Fig.~\ref{fig:N20f5}, we plot our numerical minimization results
with $f_0=5$ while all the other parameters are the same as above.
Then, the value of $Kl \approx 4.3589$ still falls between $j_{11}
\approx 3.8317$ and $j_{12} \approx 7.0156$.
If $W_{\rm ref}$ denotes the value of the Lyapunov functional
of the homogeneous solution, we see from Eq.~(\ref{eq:lhom}) that $W_{\rm ref} /
\pi l^2 = -190$. To see the minimization performance, we check a relative
difference from this value,
\begin{equation}
\Delta \equiv \frac{W_{\rm ref}-W}{W_{\rm ref}}.
\end{equation}
We first run the Metropolis algorithm from $E_{0m} = G_{0m} = 0$ with
fixing the temperature $T$ to zero. We then find two
different minima as expected: One describes a population concentration around
$r=0$, and the other shows an annular structure which is reminiscent of an ant
mill~\cite{Couzin2003}. These patterns nicely match with our
picture in Fig.~\ref{fig:scheme}(c). Especially, the concentration around $r=0$
is essentially the same pattern that we have shown in Fig.~\ref{fig:threshold}.
However, if we start from
$T=10^1$ and gradually lower the temperature down to $T \approx 10^{-3}$, a
better minimization result is achieved and it is characterized by
systematic deviations of $E_{0m}$ from zero for $m \lesssim 10$.
The small yet finite temperature $T \approx 10^{-3}$ shows us how the modes are
affected by environmental noises.
Due to the excitation of high-$m$ modes, we observe higher concentrations of
$\rho$ and $c$ around the
origin than expected from the zero-temperature case. Such coupling
between modes would not be observed if Eq.~(\ref{eq:Wp0}) was minimized term by
term. In Fig.~\ref{fig:N20f5}, we see that $E_{01}$ is considerably greater than
that of the zero-temperature result. Higher modes with $m>1$ should thus be
excited to ensure the non-negativity of $\rho$, increasing $W$. Nevertheless,
the reduction of $W$ from $m=1$ may well overtake the increment from $m>1$,
because each mode appears with a different weight in Eq.~(\ref{eq:Wp0}).
The excitation of high-$m$ modes becomes more pronounced
as we go far above $j_{11}$: For example, let us choose $f_0=10.0$ and
$\chi_0=8.0$, for which $Kl \approx 8.8882$ is greater than $j_{12} \approx
7.0156$ but lies below $j_{13} \approx 10.1735$. We observe that the
zero-temperature Metropolis algorithm ends up with one of three different minima
shown in Fig.~\ref{fig:N20f10}(a). Once again, the annealing procedure from
$T=10^1$ to $T=10^{-3}$ finds a much better result, concentrating the most of
the population around $r=0$. Note that the amplitudes $E_{0m}$ exhibit a
nontrivial structure in Fig.~\ref{fig:N20f10}(c). It actually extends to even
higher $m > 19$ if we take more modes into account in computing
Eq.~(\ref{eq:Wp0}), but those higher modes hardly affect the radius of the
aggregate in Fig.~\ref{fig:N20f10}(a).

When the distribution $\rho(r)$ is given,
the degree of aggregation can be estimated by the Shannon entropy:
\begin{equation}
S = -\int_0^{2\pi} \int_0^l \rho(r) \log \frac{\rho(r)}{\rho_{\rm const}}~~r~dr~
d\theta.
\label{eq:entropy}
\end{equation}
Figure~\ref{fig:entropy} shows $S$ as a function of $f_0$ at two different
temperatures of the Monte Carlo calculation. The other parameters are set to the
same as in Figs.~\ref{fig:threshold} and \ref{fig:N20f5}.
When $T$ is high, the system is insensitive to
$f_0$, and $S$ does not show any significant change. For low $T$, on the other
hand, it becomes clear that a jump of $S$ exists in the vicinity of $f_0 =
f_0^\ast$. Recall that the separability assumption predicts that the system
undergoes stepwise changes as $f_0$ increases, because $Kl$ has to exceed
$j_{1m}$ to excite the $m$th mode ($m=1 ,2, \ldots$). That is,
if the assumption was valid everywhere, all the higher modes would
remain suppressed unless $Kl > j_{12}$, which requires $f_0 > 12.55$.
However, our Monte Carlo
results have shown that modes tend to be coupled to each other to reduce the
free energy to a greater extent than predicted by the separability assumption.
In other words, it implies that $S$ jumps only once at $f_0^\ast$ and then
changes continuously for higher $f_0$, which is indeed the case in
Fig.~\ref{fig:entropy}.

\begin{figure}
{\centering
\includegraphics[width=0.65\textwidth, angle=270]{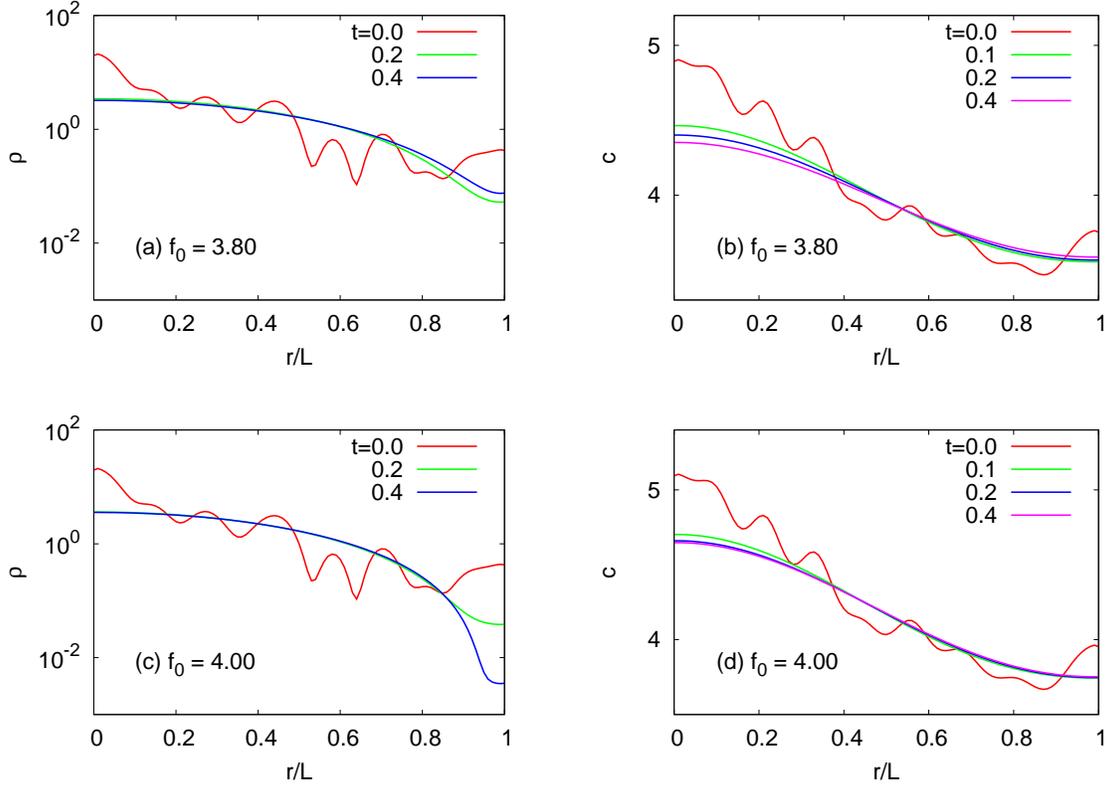}
\par}
\caption{Direct numerical simulation of Eqs.~(\ref{eq:kow1}) and (\ref{eq:kow2})
under radial symmetry
with the Forward-Time-Central-Space (FTCS) scheme. Panels (a) and (c) show
$\rho(r,t)$ and the others do $c(r,t)$. We use the same parameter as
in Figs.~\ref{fig:threshold} and \ref{fig:N20f5}, which means that the threshold
corresponds to $f_0^\ast \approx 3.92$.
Both for $f_0 = 3.80 < f_0^\ast$ (the upper panels) and $f_0 = 4.00 > f_0^\ast$
(the lower ones), the system starts from an identical configuration which is
found by the Monte Carlo calculation at some high $T$.
The only difference in the initial conditions
is the total amount of the chemical attractants because we
have set $c_{\rm const} = (f_0/g_0) \rho_{\rm const} \propto f_0$.
The time step for integration is chosen to be $\Delta t = 10^{-7}$, and the
horizontal axis is divided into $200$ grid points.
Note that the vertical axes are drawn on the log scale in panels (a) and (c) to
see the behaviour of $\rho(r,t)$ near the boundary at $r=L$.
}
\label{fig:dynamics}
\end{figure}

It is also instructive to directly consider dynamics of Eqs.~(\ref{eq:kow1}) and
(\ref{eq:kow2}) for the following reason: The idea behind our Monte Carlo
calculation is that the result can describe long-time behaviour of the real
dynamics. As mentioned in Method, the algorithm checks the non-negativity of
$\rho$ and $c$ as well as the change of $W$, so that a Monte Carlo move will be
rejected if it violates the non-negativity, even if it decreases $W$. On
the other hand, the dynamics of Eqs.~(\ref{eq:kow1}) and (\ref{eq:kow2}) does
not have such rejection but only continues with $dW/dt \le 0$
[Eq.~(\ref{eq:dW})].
Therefore, one may well ask if the dynamics always confines the system in a
physical region where both $\rho$ and $c$ are non-negative. Fortunately, the
answer is yes, as has been proved in Ref.~\citeonline{Kowalczyk2005}. We can
thus safely move on to the next question, i.e., whether the long-time behaviour
is consistent with the Monte Carlo result. Under radial symmetry, the equations
are written as
\begin{eqnarray}
\frac{\partial \rho}{\partial t} &=&
-\chi_0 \left( \frac{\rho}{r} \frac{\partial c}{\partial r} + \frac{\partial
\rho}{\partial r} \frac{\partial c}{\partial r} + \rho \frac{\partial^2
c}{\partial r^2} \right)
+ D_0 \left[ \frac{\rho}{r} \frac{\partial \rho}{\partial r} + \left(
\frac{\partial \rho}{\partial r} \right)^2 + \rho \frac{\partial^2
\rho}{\partial r^2} \right]\\
\frac{\partial c}{\partial t} &=& f_0 \rho + \nu_0 \left( \frac{1}{r}
\frac{\partial c}{\partial r} + \frac{\partial^2 c}{\partial r^2} \right)
- g_0 c.
\end{eqnarray}
We can integrate these equations numerically, e.g., with the Forward-Time
Central-Space (FTCS) method~\cite{Newman2013}, and the results are given in
Fig.~\ref{fig:dynamics}. We still use the same parameters as in
Figs.~\ref{fig:threshold} and \ref{fig:N20f5} to have a threshold at $f_0 =
f_0^\ast \approx 3.92$.
As expected, both $\rho$ and $c$ become flatter as time goes by when $f_0 <
f_0^\ast$ [Figs.~\ref{fig:dynamics}(a) and (b)]. On the other hand,
when $f_0 > f_0^\ast$ [Figs.~\ref{fig:dynamics}(c) and (d)], $\rho$ and $c$
instead converge to inhomogeneous distribution functions, respectively, which
exactly match with the ones in Fig.~\ref{fig:threshold}.
Moreover, the time evolution undergoes critical slowing down as we approach
$f_0^\ast$. It is consistent with the linear-stability analysis in which the
eigenvalue governing the mode growth (or decay) vanishes at the threshold.
However, we also note that the naive FTCS scheme becomes unstable at large $t$,
violating the non-negativity condition. This must be a numerical artefact
because, as mentioned above, the dynamics itself preserves the
non-negativity of $\rho$ and $c$~\cite{Kowalczyk2005}. A better alternative
could be to utilise the operator-splitting scheme~\cite{Lee2006}, incorporating
exact solutions of the porous-medium equation (see, e.g.,
Ref.~\citeonline{Pamuk2005}).

\section*{Summary}
\label{sec:sum}

In summary, we have investigated a variant of the Patlak-Keller-Segel model in
which pressure is assumed to increase linearly with the density of the organisms
[Eqs.~(\ref{eq:kow1}) and (\ref{eq:kow2})]. We have derived its Lyapunov
functional $W$ in Eq.~(\ref{eq:W}), which may also be called the free energy of
this system. The linear stability analysis of the homogeneous solution predicts a jump in the amplitude of aggregation as a parameter $K$, defined in
Eq.~(\ref{eq:K}), exceeds $j_{11}/l$. We have checked this transition by using
the exact Lyapunov functional, simplified for radially symmetric solutions
[Eq.~(\ref{eq:Wp0})].
The system converges to two different states depending on
in which direction the transition point is approached. At the transition point,
however, $W$ is independent of the amplitude of aggregation and a continuous
spectrum of infinitely many states exists between the two states with exactly
the same value of $W$. The transition is thus continuous.

Our numerical calculation furthermore shows that Eq.~(\ref{eq:Wp0}) has multiple
local minima (Figs.~\ref{fig:N20f5} and \ref{fig:N20f10}).
It is an open question if the existence of multiple local minima in $W$ is due
to the fact that we have restricted ourselves to radially symmetric solutions.
That is, if we relaxed the symmetry requirement, some of the local minima could
be connected to others via non-symmetric states. For example, the
annular structure in Fig.~\ref{fig:N20f5}(a) has relatively high $W$ than other
minima, and it is likely to collapse into another state in the presence of
non-symmetric perturbations. At the same time, the extended parameter space
could well introduce far more metastable states in the absence of the radial
symmetry: Reference~\citeonline{Kowalczyk2008} shows us one of such states
obtained with the finite-element method. To check those possibilities, we are
currently working with the full normal-mode expression of $W$ without the radial
symmetry.

\section*{Method}
\label{sec:method}

In minimizing a partial sum of $W$ from $m=0$ to $m = \hat{m}$
[Eq.~(\ref{eq:Wp0})] numerically, we treat the total mass $M$
[Eq.~(\ref{eq:mass})] and temperature $T$ as input parameters.
The initial state is defined by a set of variables, $G_{00} \equiv c_{\rm const}
= M/(\pi l^2)$ and $E_{01} = E_{02} = \ldots = G_{01} = G_{02} = \ldots = 0$,
from which $W$ is computed. Note that $E_{00} \equiv \rho_{\rm const} = M/(\pi
l^2)$ is a constant that will not be updated throughout the minimization
procedure. We generate a neighbouring state in the following way: We first
choose a mode $m \in [0, \ldots, \hat{m}]$. If $m>0$, we add two independent
random numbers $r_E$ and $r_G$, each of which is taken from $[-0.1,0.1)$, to the
corresponding amplitudes $E_{0m}$ and $G_{0m}$, respectively. If $m=0$, on the
other hand, only $G_{00}$ will be updated by $r_G \in [-0.1, 0.1)$ because
$E_{00} = \rho_{\rm const}$ should remain constant.
From this neighbouring state, we can calculate the Lyapunov functional, and let
us denote its value $W'$. We basically employ the standard Metropolis algorithm
to determine whether to accept the move to this
neighbouring state: We first check if the move satisfies $W' \le W$. Otherwise,
we draw a random number from $[0,1)$ and check if it is smaller than
$\exp[(W-W')/T]$.
If either of those two conditions is met, we proceed to check
if the move leaves both $\rho$ and $c$ non-negative everywhere inside the disc
by dividing the region into a sufficiently fine mesh compared to the variations
of the highest mode with $\hat{m}$. In short, we carry out the move only if it
is accepted by the Metropolis algorithm without violating the non-negativity.
One Monte Carlo step consists of $(\hat{m}+1)$ such attempts to move
to neighbouring states.

We test the algorithm by running it at $T=0$ to obtain the expected results such
as in the inset of Fig.~\ref{fig:N20f10}(a). To find a better minimum of $W$, we
choose an annealing schedule as $T = 10 \times (1.2)^{-n}$ with $n=0, 1, \ldots,
50$, and take $1.5 \times 10^4$ Monte Carlo steps at each $T$
(Figs.~\ref{fig:N20f5} and \ref{fig:N20f10}). We also note that we have added
calculations with $T=0$ at the end of this annealing schedule for clarity in
Fig.~\ref{fig:threshold}.


\begin{thebibliography}{10}
\expandafter\ifx\csname url\endcsname\relax
  \def\url#1{\texttt{#1}}\fi
\expandafter\ifx\csname urlprefix\endcsname\relax\def\urlprefix{URL }\fi
\providecommand{\bibinfo}[2]{#2}
\providecommand{\eprint}[2][]{\url{#2}}

\bibitem{Holldobler1990}
\bibinfo{author}{H\"{o}lldobler, B.} \& \bibinfo{author}{Wilson, E.~O.}
\newblock \emph{\bibinfo{title}{The Ants}} (\bibinfo{publisher}{The Belknap
  Press}, \bibinfo{address}{Cambridge}, \bibinfo{year}{1990}).

\bibitem{Vicsek2012}
\bibinfo{author}{Vicsek, T.} \& \bibinfo{author}{Zafeiris, A.}
\newblock \bibinfo{title}{Collective motion}.
\newblock \emph{\bibinfo{journal}{Physics Reports}}
  \textbf{\bibinfo{volume}{517}}, \bibinfo{pages}{71--140}
  (\bibinfo{year}{2012}).

\bibitem{VelaPerez2015}
\bibinfo{author}{Vela-P\'{e}rez, M.}, \bibinfo{author}{Fontelos, M.~A.} \&
  \bibinfo{author}{Garnier, S.}
\newblock \bibinfo{title}{From individual to collective dynamics in {A}rgentine
  ants (\emph{Linepithema humile})}.
\newblock \emph{\bibinfo{journal}{Math. Biosci.}}
  \textbf{\bibinfo{volume}{262}}, \bibinfo{pages}{56--64}
  (\bibinfo{year}{2015}).

\bibitem{Fewell1988}
\bibinfo{author}{Fewell, J.~H.}
\newblock \bibinfo{title}{Energetic and time costs of foraging in harvester
  ants, \emph{Pogonomyrmex occidentalis}}.
\newblock \emph{\bibinfo{journal}{Behav. Ecol. Sociobiol.}}
  \textbf{\bibinfo{volume}{22}}, \bibinfo{pages}{401--408}
  (\bibinfo{year}{1988}).

\bibitem{Denny2001}
\bibinfo{author}{Denny, A.~J.}, \bibinfo{author}{Wright, J.} \&
  \bibinfo{author}{Grief, B.}
\newblock \bibinfo{title}{Foraging efficiency in the wood ant, \emph{Formica
  rufa}: is time of the essence in trail following?}
\newblock \emph{\bibinfo{journal}{Anim. Behav.}} \textbf{\bibinfo{volume}{61}},
  \bibinfo{pages}{139--146} (\bibinfo{year}{2001}).

\bibitem{Couzin2003}
\bibinfo{author}{Couzin, I.~D.} \& \bibinfo{author}{Franks, N.~R.}
\newblock \bibinfo{title}{Self-organized lane formation and optimized traffic
  flow in army ants}.
\newblock \emph{\bibinfo{journal}{Proc. R. Soc. Lond. B}}
  \textbf{\bibinfo{volume}{270}}, \bibinfo{pages}{139--146}
  (\bibinfo{year}{2003}).

\bibitem{Dussutour2004}
\bibinfo{author}{Dussutour, A.}, \bibinfo{author}{Fourcassie\'{e}, V.},
  \bibinfo{author}{Helbing, D.} \& \bibinfo{author}{Deneubourg, J.-L.}
\newblock \bibinfo{title}{Optimal traffic organization in ants under crowded
  conditions}.
\newblock \emph{\bibinfo{journal}{Nature}} \textbf{\bibinfo{volume}{428}},
  \bibinfo{pages}{70--73} (\bibinfo{year}{2004}).

\bibitem{Tao2004}
\bibinfo{author}{Tao, T.}, \bibinfo{author}{Nakagawa, H.},
  \bibinfo{author}{Yamasaki, M.} \& \bibinfo{author}{Nishimori, H.}
\newblock \bibinfo{title}{Flexible foraging of ants under unsteadily varying
  environment}.
\newblock \emph{\bibinfo{journal}{J. Phys. Soc. Jpn.}}
  \textbf{\bibinfo{volume}{73}}, \bibinfo{pages}{2333--2341}
  (\bibinfo{year}{2004}).

\bibitem{Dussutour2006}
\bibinfo{author}{Dussutour, A.}, \bibinfo{author}{Nicolis, S.~C.},
  \bibinfo{author}{Deneubourg, J.-L.} \& \bibinfo{author}{Fourcassi\'{e}, V.}
\newblock \bibinfo{title}{Collective decisions in ants when foraging under
  crowded conditions}.
\newblock \emph{\bibinfo{journal}{Behav. Ecol. Sociobiol.}}
  \textbf{\bibinfo{volume}{61}}, \bibinfo{pages}{17--30}
  (\bibinfo{year}{2006}).

\bibitem{Goss1989}
\bibinfo{author}{Goss, S.}, \bibinfo{author}{Aron, S.},
  \bibinfo{author}{Deneubourg, J.-L.} \& \bibinfo{author}{Pasteels, J.~M.}
\newblock \bibinfo{title}{Self-organized shortcuts in the {Argentine} ants}.
\newblock \emph{\bibinfo{journal}{Naturwissenschaften}}
  \textbf{\bibinfo{volume}{76}}, \bibinfo{pages}{579--582}
  (\bibinfo{year}{1989}).

\bibitem{Reid2010}
\bibinfo{author}{Reid, C.~R.}, \bibinfo{author}{Sumpter, D. J.~T.} \&
  \bibinfo{author}{Beekman, M.}
\newblock \bibinfo{title}{Optimisation in a natural system: {Argentine} ants
  solve the {Tower of Hanoi}}.
\newblock \emph{\bibinfo{journal}{J. Exp. Biol.}}
  \textbf{\bibinfo{volume}{214}}, \bibinfo{pages}{50--58}
  (\bibinfo{year}{2010}).

\bibitem{Oettler2013}
\bibinfo{author}{Oettler, J.} \emph{et~al.}
\newblock \bibinfo{title}{Fermat's principle of least time predicts refraction
  of ant trails at substrate borders}.
\newblock \emph{\bibinfo{journal}{PLoS ONE}} \textbf{\bibinfo{volume}{8}},
  \bibinfo{pages}{1--7} (\bibinfo{year}{2013}).

\bibitem{Vicsek1995}
\bibinfo{author}{Vicsek, T.}
\newblock \bibinfo{title}{Novel type of phase transition in a system of
  self-driven particles}.
\newblock \emph{\bibinfo{journal}{Phys. Rev. Lett.}}
  \textbf{\bibinfo{volume}{75}}, \bibinfo{pages}{1226--1229}
  (\bibinfo{year}{1995}).

\bibitem{Dorigo2004}
\bibinfo{author}{Dorigo, M.} \& \bibinfo{author}{St\"{u}tzle, T.}
\newblock \emph{\bibinfo{title}{Ant Colony Optimization}}
  (\bibinfo{publisher}{A Bradford Book}, \bibinfo{address}{London},
  \bibinfo{year}{2004}).

\bibitem{Jafar2010}
\bibinfo{author}{{Mohamed~Jafar}, O.~A.} \& \bibinfo{author}{Sivakumar, R.}
\newblock \bibinfo{title}{Ant-based clustering algorithms: A brief survey}.
\newblock \emph{\bibinfo{journal}{Int. J. Comput. Theor. Eng.}}
  \textbf{\bibinfo{volume}{2}}, \bibinfo{pages}{1793--8201}
  (\bibinfo{year}{2010}).

\bibitem{Marchetti2013}
\bibinfo{author}{Marchetti, M.~C.} \emph{et~al.}
\newblock \bibinfo{title}{Hydrodynamics of soft active matter}.
\newblock \emph{\bibinfo{journal}{Rev. Mod. Phys.}}
  \textbf{\bibinfo{volume}{85}}, \bibinfo{pages}{1143} (\bibinfo{year}{2013}).

\bibitem{Deneubourg2002}
\bibinfo{author}{Deneubourg, J.~L.}, \bibinfo{author}{Lioni, A.} \&
  \bibinfo{author}{Detrain, C.}
\newblock \bibinfo{title}{Dynamics of aggregation and emergence of
  cooperation}.
\newblock \emph{\bibinfo{journal}{Biol. Bull.}} \textbf{\bibinfo{volume}{202}},
  \bibinfo{pages}{262--267} (\bibinfo{year}{2002}).

\bibitem{Tennenbaum2015}
\bibinfo{author}{Tennenbaum, M.}, \bibinfo{author}{Liu, Z.},
  \bibinfo{author}{Hu, D.} \& \bibinfo{author}{Fernandez-Nieves, A.}
\newblock \bibinfo{title}{Mechanics of fire ant aggregations}.
\newblock \emph{\bibinfo{journal}{Nat. Mater.}} \textbf{\bibinfo{volume}{15}},
  \bibinfo{pages}{54--59} (\bibinfo{year}{2015}).

\bibitem{Patlak1953}
\bibinfo{author}{Patlak, C.~S.}
\newblock \bibinfo{title}{Random walk with persistence and external bias}.
\newblock \emph{\bibinfo{journal}{Bull. Math. Biophys.}}
  \textbf{\bibinfo{volume}{15}}, \bibinfo{pages}{311--338}
  (\bibinfo{year}{1953}).

\bibitem{Keller1970}
\bibinfo{author}{Keller, E.~F.} \& \bibinfo{author}{Segel, L.~A.}
\newblock \bibinfo{title}{Initiation of slime mold aggregation viewed as an
  instability}.
\newblock \emph{\bibinfo{journal}{J. Theor. Biol.}}
  \textbf{\bibinfo{volume}{26}}, \bibinfo{pages}{399--415}
  (\bibinfo{year}{1970}).

\bibitem{Horstmann2003}
\bibinfo{author}{Horstmann, D.}
\newblock \bibinfo{title}{From 1970 until now: the {Keller-Segel} model in
  chemotaxis and its consequences {I}}.
\newblock \emph{\bibinfo{journal}{Jahresber. Dtsch. Math. Ver.}}
  \textbf{\bibinfo{volume}{105}}, \bibinfo{pages}{103--165}
  (\bibinfo{year}{2003}).

\bibitem{Horstmann2004}
\bibinfo{author}{Horstmann, D.}
\newblock \bibinfo{title}{From 1970 until now: the {Keller-Segel} model in
  chemotaxis and its consequences {II}}.
\newblock \emph{\bibinfo{journal}{Jahresber. Dtsch. Math. Ver.}}
  \textbf{\bibinfo{volume}{106}}, \bibinfo{pages}{51--69}
  (\bibinfo{year}{2004}).

\bibitem{Biler1999}
\bibinfo{author}{Biler, P.}
\newblock \bibinfo{title}{Global solutions to some parabolic-elliptic systems
  of chemotaxis}.
\newblock \emph{\bibinfo{journal}{Adv. Math. Sci. Appl.}}
  \textbf{\bibinfo{volume}{9}}, \bibinfo{pages}{347--359}
  (\bibinfo{year}{1999}).

\bibitem{Hillen2001}
\bibinfo{author}{Hillen, T.} \& \bibinfo{author}{Painter, K.}
\newblock \bibinfo{title}{Global existence for a parabolic chemotaxis model
  with prevention of overcrowding}.
\newblock \emph{\bibinfo{journal}{Adv. Appl. Math.}}
  \textbf{\bibinfo{volume}{26}}, \bibinfo{pages}{280--301}
  (\bibinfo{year}{2001}).

\bibitem{Painter2002}
\bibinfo{author}{Painter, K.~J.} \& \bibinfo{author}{Hillen, T.}
\newblock \bibinfo{title}{Volume-filling and quorum-sensing in models for
  chemosensitive movement}.
\newblock \emph{\bibinfo{journal}{Can. Appl. Math. Quart.}}
  \textbf{\bibinfo{volume}{10}}, \bibinfo{pages}{501--543}
  (\bibinfo{year}{2002}).

\bibitem{Potapov2005}
\bibinfo{author}{Potapov, A.~B.} \& \bibinfo{author}{Hillen, T.}
\newblock \bibinfo{title}{Metastability in chemotaxis models}.
\newblock \emph{\bibinfo{journal}{J. Dyn. Differ. Equ.}}
  \textbf{\bibinfo{volume}{17}}, \bibinfo{pages}{293--329}
  (\bibinfo{year}{2005}).

\bibitem{Kowalczyk2005}
\bibinfo{author}{Kowalczyk, R.}
\newblock \bibinfo{title}{Preventing blow-up in a chemotaxis model}.
\newblock \emph{\bibinfo{journal}{J. Math. Anal. Appl.}}
  \textbf{\bibinfo{volume}{305}}, \bibinfo{pages}{566--588}
  (\bibinfo{year}{2005}).

\bibitem{Kowalczyk2008}
\bibinfo{author}{Kowalczyk, R.} \& \bibinfo{author}{Szyma\'{n}ska, Z.}
\newblock \bibinfo{title}{On the global existence of solutions to an
  aggregation model}.
\newblock \emph{\bibinfo{journal}{J. Math. Anal. Appl.}}
  \textbf{\bibinfo{volume}{343}}, \bibinfo{pages}{379--398}
  (\bibinfo{year}{2008}).

\bibitem{Tello2007}
\bibinfo{author}{{Ignacio Tello}, J.} \& \bibinfo{author}{Winkler, M.}
\newblock \bibinfo{title}{A chemotaxis system with logistic source}.
\newblock \emph{\bibinfo{journal}{Commun. Part. Diff. Eq.}}
  \textbf{\bibinfo{volume}{32}}, \bibinfo{pages}{849--877}
  (\bibinfo{year}{2007}).

\bibitem{Hittmeir2011}
\bibinfo{author}{Hittmeir, S.} \& \bibinfo{author}{J{\"u}ngel, A.}
\newblock \bibinfo{title}{Cross diffusion preventing blow-up in the
  two-dimensional {Keller–Segel} model}.
\newblock \emph{\bibinfo{journal}{SIAM J. Math. Anal.}}
  \textbf{\bibinfo{volume}{43}}, \bibinfo{pages}{997--1022}
  (\bibinfo{year}{2011}).

\bibitem{Bedrossian2016}
\bibinfo{author}{Bedrossian, J.} \& \bibinfo{author}{He, S.}
\newblock \bibinfo{title}{Suppression of blow-up in {Patlak-Keller-Segel} via
  shear flows}.
\newblock \bibinfo{note}{{arXiv}:1609.02866}.

\bibitem{Hillen2009}
\bibinfo{author}{Hillen, T.} \& \bibinfo{author}{Painter, K.~J.}
\newblock \bibinfo{title}{A user's guide to {PDE} models for chemotaxis}.
\newblock \emph{\bibinfo{journal}{J. Math. Biol.}}
  \textbf{\bibinfo{volume}{58}}, \bibinfo{pages}{183--217}
  (\bibinfo{year}{2009}).

\bibitem{Amorim2015}
\bibinfo{author}{Amorim, P.}
\newblock \bibinfo{title}{Modeling ant foraging: A chemotaxis approach with
  pheromones and trail formation}.
\newblock \emph{\bibinfo{journal}{J. Theor. Biol.}}
  \textbf{\bibinfo{volume}{385}}, \bibinfo{pages}{160--173}
  (\bibinfo{year}{2015}).

\bibitem{Strogatz2001}
\bibinfo{author}{Strogatz, S.~H.}
\newblock \emph{\bibinfo{title}{Nonlinear Dynamics and Chaos: With Applications
  to Physics, Biology, Chemistry, and Engineering}}
  (\bibinfo{publisher}{Westview Press}, \bibinfo{address}{Boulder, CO},
  \bibinfo{year}{2001}).

\bibitem{Fatkullin2013}
\bibinfo{author}{Fatkullin, I.}
\newblock \bibinfo{title}{A study of blow-ups in the {Keller-Segel} model of
  chemotaxis}.
\newblock \emph{\bibinfo{journal}{Nonlinearity}} \textbf{\bibinfo{volume}{26}},
  \bibinfo{pages}{81--94} (\bibinfo{year}{2013}).

\bibitem{Edwards1972}
\bibinfo{author}{Edwards, J.~T.} \& \bibinfo{author}{Thouless, D.~J.}
\newblock \bibinfo{title}{Numerical studies of localization in disordered
  systems}.
\newblock \emph{\bibinfo{journal}{J. Phys. C}} \textbf{\bibinfo{volume}{5}},
  \bibinfo{pages}{807--820} (\bibinfo{year}{1972}).

\bibitem{Calvez2008}
\bibinfo{author}{Calvez, V.} \& \bibinfo{author}{Corrias, L.}
\newblock \bibinfo{title}{The parabolic-parabolic {Keller-Segel} model in
  ${R}^2$}.
\newblock \emph{\bibinfo{journal}{Commun. Math. Sci.}}
  \textbf{\bibinfo{volume}{6}}, \bibinfo{pages}{417--447}
  (\bibinfo{year}{2008}).

\bibitem{Tsallis1988}
\bibinfo{author}{Tsallis, C.}
\newblock \bibinfo{title}{Possible generalization of {Boltzmann-Gibbs}
  statistics}.
\newblock \emph{\bibinfo{journal}{J. Stat. Phys.}}
  \textbf{\bibinfo{volume}{52}}, \bibinfo{pages}{479--487}
  (\bibinfo{year}{1988}).

\bibitem{Childress1981}
\bibinfo{author}{Childress, S.} \& \bibinfo{author}{Percus, J.~K.}
\newblock \bibinfo{title}{Nonlinear aspects of chemotaxis}.
\newblock \emph{\bibinfo{journal}{Math. Biosci.}}
  \textbf{\bibinfo{volume}{56}}, \bibinfo{pages}{217--237}
  (\bibinfo{year}{1981}).

\bibitem{Regnier1968}
\bibinfo{author}{Regnier, F.~E.} \& \bibinfo{author}{Law, J.~H.}
\newblock \bibinfo{title}{Insect pheromones}.
\newblock \emph{\bibinfo{journal}{J. Lipid Res.}} \textbf{\bibinfo{volume}{9}},
  \bibinfo{pages}{541--551} (\bibinfo{year}{1968}).

\bibitem{Boas2006}
\bibinfo{author}{Boas, M.~L.}
\newblock \emph{\bibinfo{title}{Mathematical Methods in the Physical Sciences}}
  (\bibinfo{publisher}{Wiley}, \bibinfo{address}{Hoboken, NJ},
  \bibinfo{year}{2006}), \bibinfo{edition}{3} edn.

\bibitem{Newman2013}
\bibinfo{author}{Newman, M. E.~J.}
\newblock \emph{\bibinfo{title}{Computational Physics}}
  (\bibinfo{publisher}{CreateSpace Independent}, \bibinfo{address}{United
  States}, \bibinfo{year}{2013}).

\bibitem{Lee2006}
\bibinfo{author}{Lee, H.~K.}, \bibinfo{author}{Kown, C.} \&
  \bibinfo{author}{Park, H.}
\newblock \bibinfo{title}{Equivalence of operator-splitting schemes for the
  integration of the {Langevin} equation}.
\newblock \emph{\bibinfo{journal}{J. Stat. Mech.: Theory Exp.}}
  \textbf{\bibinfo{volume}{2006}}, \bibinfo{pages}{P08021}
  (\bibinfo{year}{2006}).

\bibitem{Pamuk2005}
\bibinfo{author}{Pamuk, S.}
\newblock \bibinfo{title}{Solution of the porous media equation by {Admomian}'s
  decomposition method}.
\newblock \emph{\bibinfo{journal}{Phys. Lett. A}}
  \textbf{\bibinfo{volume}{344}}, \bibinfo{pages}{184--188}
  (\bibinfo{year}{2005}).

\end{thebibliography}

\section*{Acknowledgments}
We gratefully acknowledge discussions with Su Do Yi and Bertrand R\"ohner.
S.K.B. and B.J.K. were supported by Basic Science Research Program through the
National Research Foundation of Korea (NRF) funded by the Ministry of Science,
ICT and Future Planning with grant No. NRF-2017R1A1A1A05001482 and
NRF-2017R1A2B2005957, respectively.

\section*{Author contributions statement}
B.J.K. conceived the study. S.K.B. carried out the calculations.
S.K.B. and B.J.K. wrote the paper.

\section*{Additional information}

Competing financial interests: The authors declare no competing financial interests.

\end{document}